\documentclass[preprint,showpacs,showkeys,floats,floatfix]{revtex4}
 
\usepackage{graphicx}
\usepackage{amsmath}
\usepackage{bm}
\def\to{\widetilde{\Omega}}
\begin{document}

\title{Comparative study of quantum anharmonic potentials}

\author{Paolo Amore}
\email{paolo@ucol.mx}
\author{Alfredo Aranda}
%\email{fefo@ucol.mx}
\affiliation{Facultad de Ciencias, Universidad de Colima, \\
  Bernal D\'{\i}az del Castillo 340, Colima, Colima, M\'exico} 

\author{Arturo De Pace}
%\email{depace@to.infn.it}
\affiliation{Istituto Nazionale di Fisica Nucleare, sezione di Torino, \\
  via P. Giuria 1, I-10125, Torino, Italy}

\author{Jorge A. L\'opez}
%\email{jorgelopez@utep.edu}
\affiliation{Physics Department, University of Texas at El Paso, \\
  El Paso, Texas, USA}

\begin{abstract}
We perform a study of various anharmonic potentials using a recently developed 
method. We calculate both the wave functions and the energy eigenvalues 
for the ground and first excited states of the quartic, sextic and octic
potentials with high precision, comparing the results with other techniques
available in the literature.
\end{abstract}


\maketitle

%%%%%%%%%%%%%%%%%%%%%%%%%%%%%%%%%%%%%%%%%%%%%%%%%%

\section{Introduction}

Ordinary perturbation theory often leads to asymptotically divergent series
\cite{Gui90}. Several techniques have been devised in the past in order to
improve the convergence of the standard perturbative expansion (see, e.~g.,
Refs.~\cite{Hatsuda:1996vp,BB96,Gui95,Jan95,Yuk91} and references therein).

Some of the ``optimized expansions'' that have been proposed are based upon the
so-called linear delta expansion (LDE) \cite{lde,lde1}, where the terms in the
Schr\"odinger equation are rearranged in ``non-perturbative'' and
``perturbative'' pieces, with the contributions of the latter being minimized
by introducing a suitably chosen variational parameter, that is by applying the
principle of minimal sensitivity (PMS) \cite{Ste81}.

Recently, in Ref.~\cite{AAD03}, an improved ansatz for the choice of the
parameter that allows for an ``optimal'' expansion has been proposed and tested
in the case of the ground and first excited states of the quantum (quartic)
anharmonic oscillator (AHO). Within this scheme calculating an observable with
higher precision --- that is going to higher order in the expansion ---
requires the solution of algebraic equations of increasing order.
This can be done analytically, thus allowing one to get analytical expressions 
for both the energies and the wave functions of the AHO at any desired order.

Given the success of the method for the quartic AHO, one would like to have a
formal proof of the convergence of the method and to verify whether  it works
for more general potentials.
In this paper we address this latter issue, extending the previous study to the
sextic and octic AHO potentials: This is of interest for the study of realistic
potentials --- since they can often be approximated as polynomials --- and also
in itself, since general anharmonic potentials have applications in studies of 
nonlinear mechanics, molecular physics, quantum optics, nuclear physics and 
field theory \cite{PM02,Meurice02}. 

We also improve the convergence of the method for the calculation of the energy
with respect to Ref.~\cite{AAD03}, by introducing a variational principle, that
is by considering, at a given order, not the expression for the energy stemming
from the LDE, but the matrix element of the hamiltonian obtained using the wave
function at that order and applying the PMS to it: Since the resulting
expression is quadratic in the wave function, one gets higher order
contributions that turn out to substantially improve the convergence of the
expansion and, at the same time, to satisfy a variational principle.
 
The paper is organized in the following way: Section~\ref{sec_met} contains a
description of the method and a discussion of the computation of the energy
through its expectation value and its relation to the variational principle. 
In Section~\ref{sec_res} we present our results for the quartic, sextic and
octic potentials, comparing them with those obtained with other techniques
available in the literature. Finally, our conclusions are presented in
Section~\ref{sec_conc}. 

\section{The method}
\label{sec_met}

The method employed in this paper has been originally devised in 
Ref.~\cite{AAD03} and applied to the study of the quartic anharmonic
potential. It relies on the the identification of three different scales in the
problem, which reflect in a different behavior of the wave function: 
An asymptotic scale, which is fully determined by the potential; an
intermediate scale, where the wave function decays exponentially, although with
less strength; a short distance scale where the wave function is sizable.

In this work we consider the application of our method to anharmonic potentials
of the form $V(x) = m\omega^2 x^2/2+\mu/(2 N) \ x^{2 N}$ and discuss the accuracy of the
approximations obtained in this framework in the case of quartic, sextic and
octic potentials (with $N=2,3,4$ respectively). 

The Schr\"odinger equation in the present case reads
\begin{equation}
  \left[ - \frac{\hbar^2}{2m} \frac{d^2}{d x^2} +  \frac{m
    \omega^2}{2} x^2 + \frac{\mu}{2 N} \ x^{2 N} \right] \psi_n(x) = E_n
    \psi_n(x), 
\label{eq2}
\end{equation}
where $\mu$ is the anharmonic coupling, $\psi_n(x)$ is the wave function of the
$nth$ excited state and $E_n$ its energy.

Although one cannot find the solution of Eq.~(\ref{eq2}) exactly,  it is
possible to determine the asymptotic behavior of $\psi_n(x)$ in the region of
large $x$  ($x \rightarrow \infty$) by substituting the ansatz 
$\psi_n(x) \propto e^{- \gamma |x|^p}$ into Eq.~(\ref{eq2}). 
One obtains $p=N+1$ and $\gamma = \sqrt{\mu m/N}/((N+1) \hbar)$.  

In the spirit of Ref.~\cite{AAD03} we therefore write the wave function as
\begin{equation}
  \psi_n(x) =   e^{- \gamma |x|^{N+1} - \beta x^2 } \xi_n(x),
\label{eq3}
\end{equation}
where the exponential takes care of the correct behavior in the limit 
$|x| \rightarrow \infty$. Notice that the quadratic term in the exponential 
does not affect the behavior at large distances, but is relevant at
intermediate scales. 
The coefficient $\beta = m \ \to/2\hbar$ is written in terms of the frequency 
$\to\equiv\sqrt{\omega^2+\Omega^2}$ ($\Omega$ is an arbitrary parameter
introduced by hand, see below). 
The reduced wave function $\xi_n$ is well-behaved and  fulfills the 
equation\footnote{Given the symmetry properties of the wave function, we are
considering only the region $x>0$.}:
\begin{eqnarray}
  \xi_n''(x) &-& 2 \left( \frac{m \to}{\hbar} x + \sqrt{\frac{\mu m}{N}} 
    \frac{x^N}{\hbar} \right) \xi_n'(x) + \left[ \left( 2 \frac{m E_n}{\hbar^2}
    - \frac{m \to}{\hbar} \right) \right. \nonumber \\ 
  &+& \left.  \frac{m^2 \Omega^2}{\hbar^2} x^2 + \sqrt{\frac{\mu m}{N}} 
    \frac{x^{N-1}}{\hbar^2} \left(2 m \to x^2 - \hbar N \right) \right] 
    \xi_n(x) = 0 .
\label{eq4}
\end{eqnarray}
We observe that no approximation has been invoked in the derivation of
Eq.~(\ref{eq4}). 
In the spirit of the Linear Delta Expansion (LDE) we now write Eq.~(\ref{eq4})
as 
\begin{eqnarray}
  \xi_n''(x) &-& \left[ \frac{2 m \to x}{\hbar} \right] \xi'_n(x) + 
    \left[\frac{2 m E_n}{\hbar^2} - \frac{m \to }{\hbar} \right] \xi_n(x) 
    = \delta \left\{ 2 \sqrt{\frac{\mu m}{N}} \frac{x^N}{\hbar} \xi_n'(x) 
    \right. \nonumber \\ 
  &-& \left. \frac{m^2 \Omega^2}{\hbar^2} x^2 \xi_n(x) - \sqrt{\frac{\mu m}{N}}
    \frac{x^{N-1}}{\hbar^2} \left(2 m \to x^2 - \hbar N \right) \xi_n(x)
    \right\} .
\label{eq6}
\end{eqnarray}
In writing this equation we have added a parameter $\delta$, which was not
present in  Eq.~(\ref{eq4}). 
For $\delta=1$ one recovers the original equation in Eq.~(\ref{eq4}), while by 
taking $\delta = 0$ one obtains the equation for the Hermite polynomials,
corresponding to a harmonic oscillator of frequency $\to$. Notice that $\delta$
here is used as a power counting device: As a matter of fact we will treat the 
right hand side of Eq.~(\ref{eq6}) as a perturbation, although its size is not
necessarily small, given the arbitrary nature of the parameter $\Omega$. 
An optimal choice of $\Omega$ will make the right hand side of Eq.~(\ref{eq6})
small. 

We now write the following expansions:
\begin{equation}
  \xi_n(x) = \sum_{j=0}^\infty \delta^j \xi_{nj}(x), \ \ \ E_n =
    \sum_{j=0}^\infty \delta^j E_{nj}
\label{eq6a}
\end{equation}
and substitute them in Eq.~(\ref{eq6}), thus generating a hierarchy of
equations, corresponding to the different orders in $\delta$. Such equations,
which take the form of the equation for the Hermite polynomials in the presence
of a source term, can be solved sequentially up to some finite order. 

\begin{figure}[t]
\begin{center}
\includegraphics[width=0.95\textwidth]{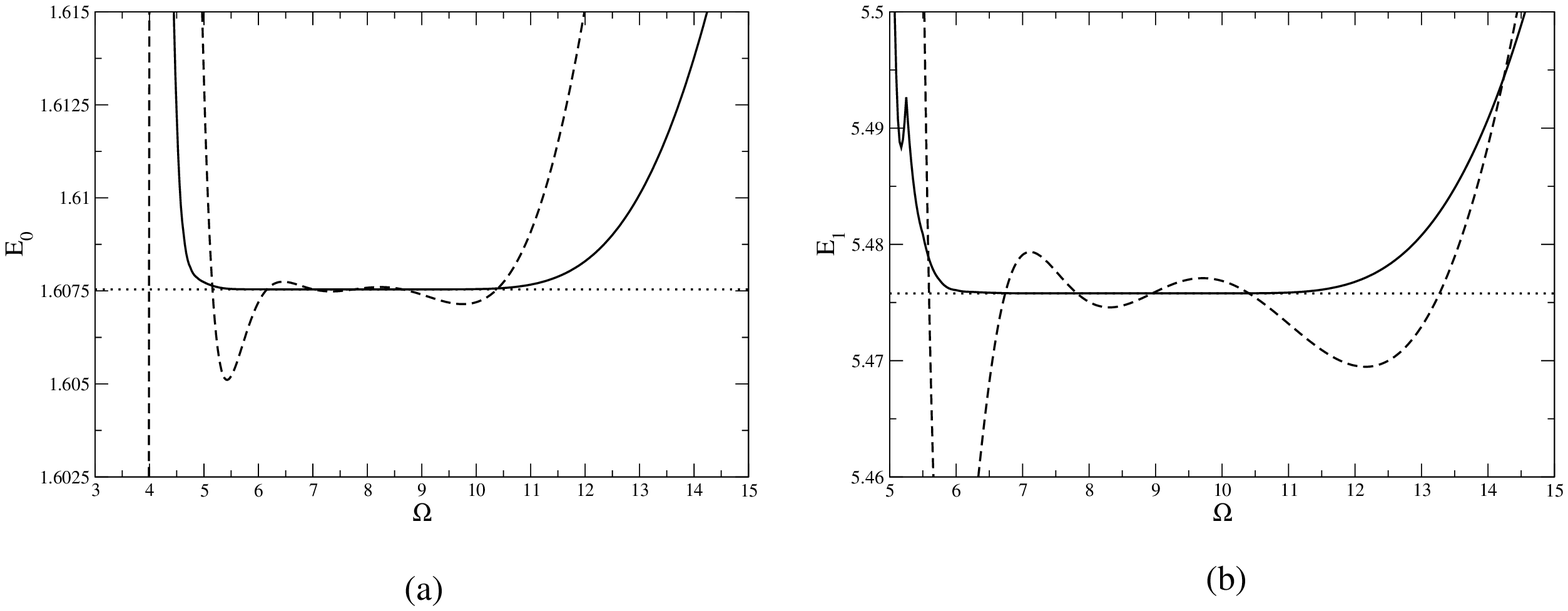}
\caption{Comparison between the ground state (a) and first excited state (b) 
  energy of the quartic oscillator calculated to
  order $15$ directly through the expansion of Eq.~(\protect\ref{eq6a})
  (dashed) and the one calculated as $\langle\psi_i|H|\psi_i\rangle$, with
  $\psi_i$ at the same order ($i=0,1$), as a function of the arbitrary
  parameter $\Omega$. The dotted line is the result of Ref.~\cite{mei97}.
  We use $m=1/2$, $\omega = 2$ and $\mu = 8$.}
\label{fig:compsandw}
\vskip 1cm
\includegraphics[width=0.95\textwidth]{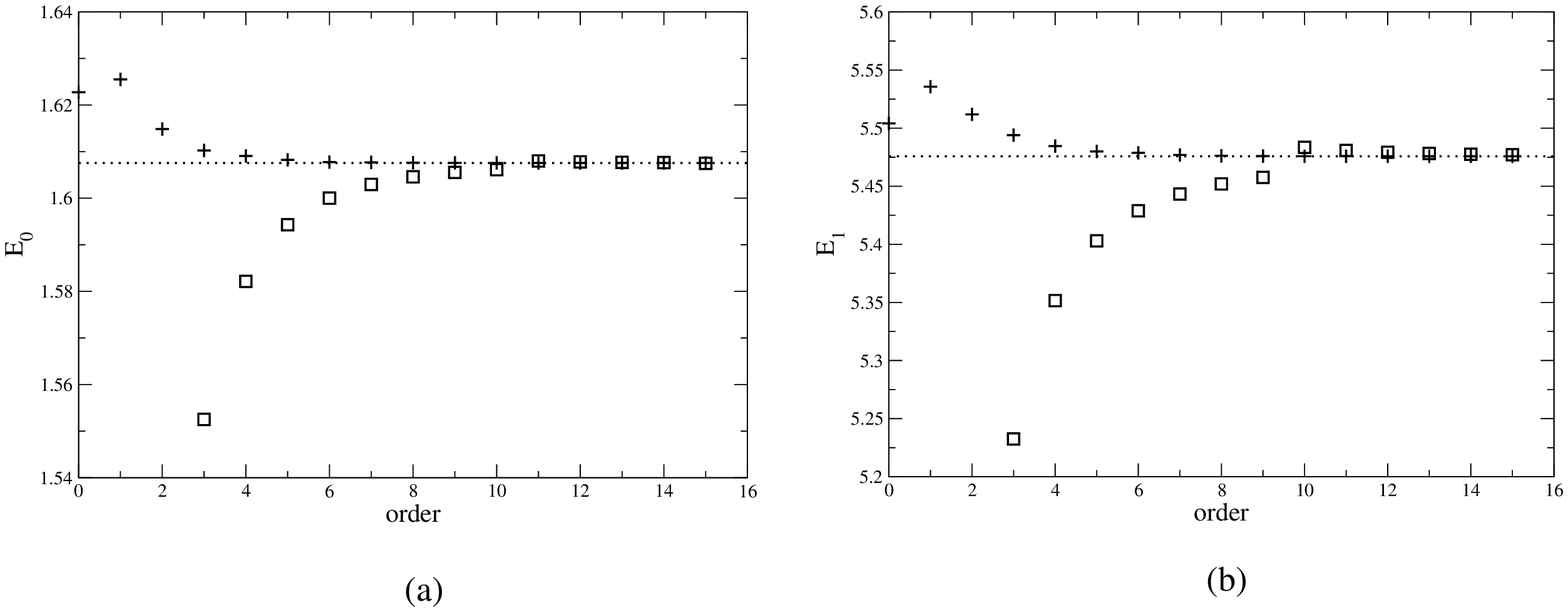}
\caption{Comparison between the ground state (a) and first excited state (b) 
  energy of the quartic oscillator calculated through the expansion of the
  energy (squares) and the one calculated as $\langle\psi_i|H|\psi_i\rangle$
  (crosses). The dotted line is the result of Ref.~\cite{mei97}. 
  We use $m=1/2$, $\omega = 2$ and $\mu = 8$.}
\label{fig:compsandwene}
\end{center}
\end{figure}
Given the perturbative nature of the approach that we are using, all the
results, obtained to a finite order in perturbation theory, will display a
dependence upon the arbitrary frequency $\Omega$. However this dependence is
artificial since the solution of Eq.~(\ref{eq6}) does not depend upon
$\Omega$. In the framework of the LDE such dependence is minimized by
applying the Principle of Minimal Sensitivity (PMS) \cite{Ste81}, i.e. by
requiring that a given observable $O$ (the energy, for example)  be locally
independent of $\Omega$: 
\begin{equation}
  \frac{\partial O}{\partial \Omega} = 0 .
\end{equation}
In the present work we will enforce the PMS by using the energy of the state as
the observable. We need however to make an important point: By solving the
equations corresponding to the different perturbative orders, we obtain two
different estimates of the energy of the solution. One corresponds to the
expansion in Eq.~(\ref{eq6a}), which was indeed used in \cite{AAD03}; the
other corresponds to calculating the energy as the expectation value of the
Hamiltonian in the state described by the wave function obtained in
Eq.~(\ref{eq6a}). This second choice turns out to provide a much better
estimate of the energy: As a matter of fact, the energy calculated in this way
not only contains contributions of higher order in $\delta$ (the  expectation
value of the Hamiltonian is bilinear in the wave function) but it is also
constrained by the variational principle (at least for the ground state) to lie
above the exact result. Therefore the PMS applied to the expectation value of
the Hamiltonian corresponds, for the ground state, to the statement of the
variational principle. 

This finding is illustrated in Fig.~\ref{fig:compsandw}, where we compare the
energy of the ground and first excited states of the quartic AHO --- evaluated
at order 15 --- to the average value of the Hamiltonian --- obtained using the
wave function at the same order --- as a function of $\Omega$.

In Fig.~\ref{fig:compsandwene}, on the other hand, one can see the ground and
first excited state energies of the quartic AHO as a function of the
perturbative order $m$, obtained after application of the PMS both to
$E^{(m)}_n$ of Eq.~(\ref{eq6a}) and to
$\langle\psi_i^{(m)}|H|\psi_i^{(m)}\rangle$.  The faster convergence to 
the correct result in the latter case is quite apparent.

It is interesting that also in the case of the excited state the approach based
upon the minimization of the average value of the Hamiltonian seems to satisfy
a variational principle. 

\section{Results}
\label{sec_res}

Let us start by comparing the accuracy of our method with the one of
Ref.~\cite{Bellet:1994mf} which is based on a variational improvement of
the ordinary perturbation theory that gives a convergent sequence of approximations.

\begin{figure}[t]
\begin{center}
\includegraphics[width=8.5cm]{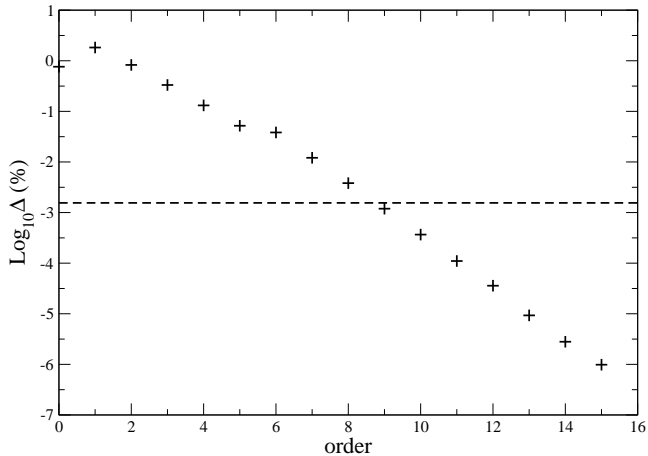}
\caption{Logarithm of the error for the ground state energy of the
 quartic oscillator as a function of the order. The dashed line is the error of
 Eq.~(11) of Ref.~\protect\cite{Bellet:1994mf}, obtained  after $47$
 iterations. We use $\omega = 0$ and $\mu = 1$.}
\label{fig:bellet}
\end{center}
\end{figure}
In Fig.~\ref{fig:bellet} we display the logarithm of the percentile error 
for the ground state energy of the quartic oscillator as a
function of the approximation order. The error is defined as
\begin{equation}
\label{eq:error}
  \Delta = 100 \times \left|\frac{E_n^{(m)}-E_n}{E_n} \right|,
\end{equation}
where $E_n$ is the ``exact'' value of the energy of the $n$-th state and
$E_n^{(m)}$ the approximate estimation to order $m$.
The results using the present method (crosses) are compared to the outcome of
the calculation in Ref.~\cite{Bellet:1994mf}, obtained after 47 iterations, for
the ground state of the quartic AHO (the ``exact'' value of $E_0$ is taken from Ref.~\cite{Janke:1995wt}).

We now turn to compare our results with those of Ref.~\cite{mei97}, where the
quartic, sextic and octic AHO have been thoroughly analyzed with a method based
on the generalized Bloch equation. This method calculates iteratively certain matrix
elements of the wave operator (wave function expansion coefficients) and then uses
a renormalization technique.

\begin{figure}[p]
\begin{center}
\includegraphics[width=0.95\textwidth]{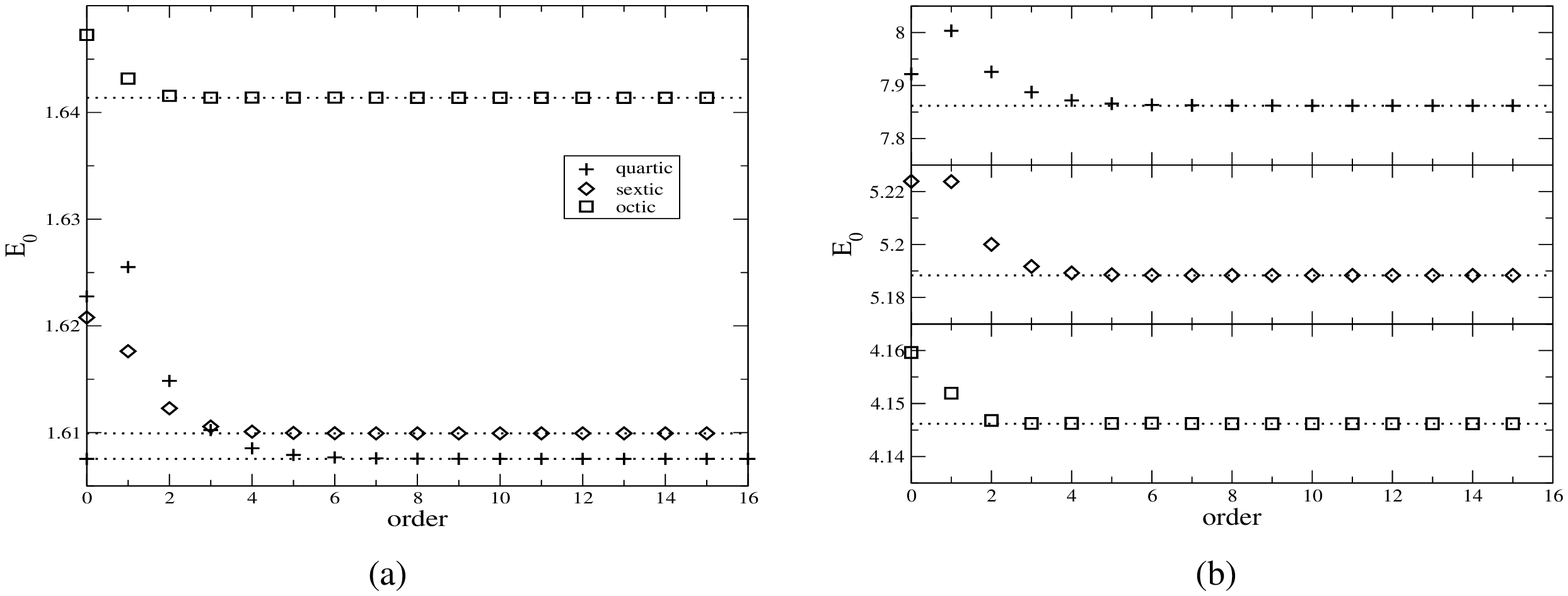}
\caption{Energy of the ground state of the quartic, sextic and octic anharmonic
  oscillators, as a function of the approximation order, compared with the
  results of Ref.~\protect\cite{mei97} (dotted). 
  We assume $m=1/2$ and $\omega=2$; $\mu= (8,12,16)$, respectively, in panel (a)
  and $\mu= (1600,2400,3200)$, respectively, in panel (b).} 
\label{fig:E0meis}
\vskip 3cm
\includegraphics[width=0.95\textwidth]{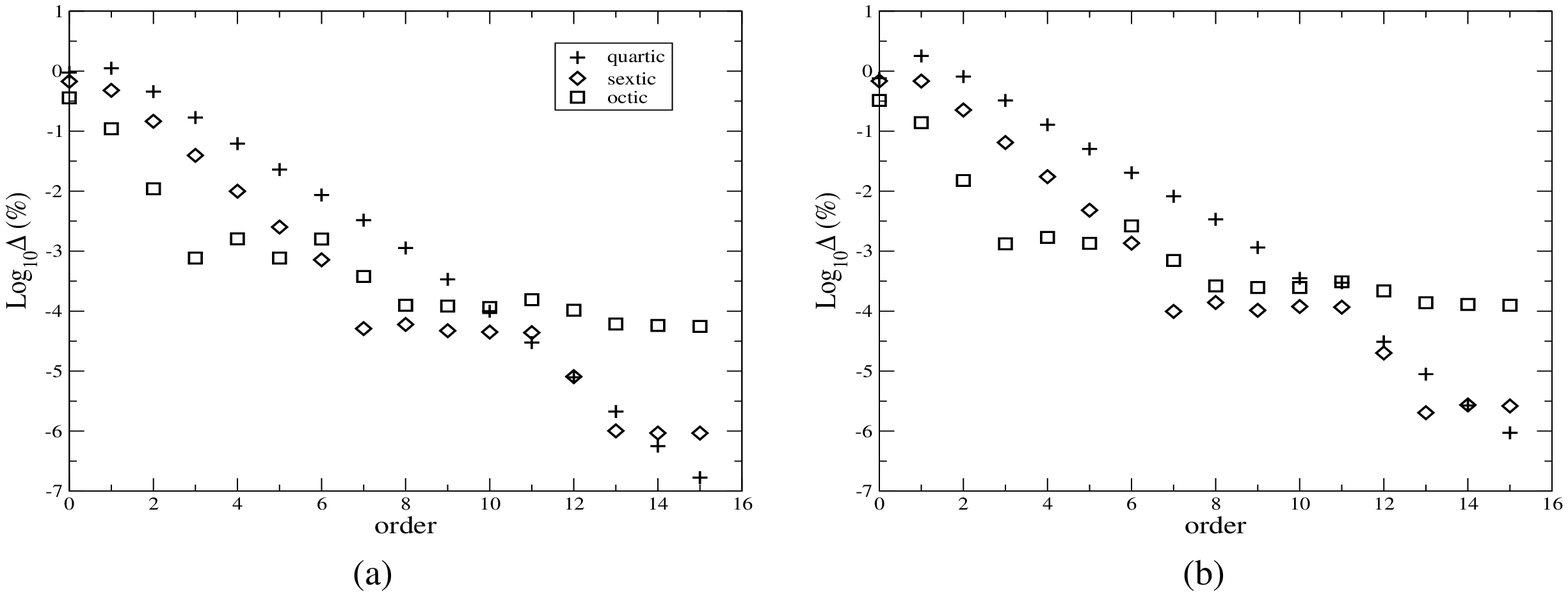}
\caption{Logarithm of the error, with respect to the results of
  Ref.~\protect\cite{mei97}, of the energy of the ground state of the quartic,
  sextic and octic anharmonic oscillators, as a function of the approximation
  order. We assume $m=1/2$ and $\omega=2$; $\mu= (8,12,16)$, respectively, in
  panel (a) and $\mu= (1600,2400,3200)$, respectively, in panel (b).}
\label{fig:D0meis}
\end{center}
\end{figure}
In Fig.~\ref{fig:E0meis} we display the energy of the ground state of the
quartic, sextic and octic AHO, as a function of the approximation order,
compared with the results of Ref.~\protect\cite{mei97}.
We have chosen the parameters of the AHO in such a way to correspond to the
cases $\beta=2$ (panel (a)) and $\beta=400$ (panel (b)).
Note that the values of Ref.~\cite{mei97} have been obtained using 100 basis
functions and several tens or hundreds of iterations.

In Fig.~\ref{fig:D0meis} we report, for the same cases of
Fig.~\ref{fig:E0meis}, the percentile error, defined as in
Eq.~(\ref{eq:error}), but taking now for $E_n$ the value calculated in
Ref.~\cite{mei97}. In other words, in Fig.~\ref{fig:D0meis} one can see, order
by order, the discrepancy of our results from those of Ref.~\cite{mei97}. 
It is apparent that even for relatively low orders the agreement is quite
good.

\begin{figure}[p]
\begin{center}
\includegraphics[width=0.95\textwidth]{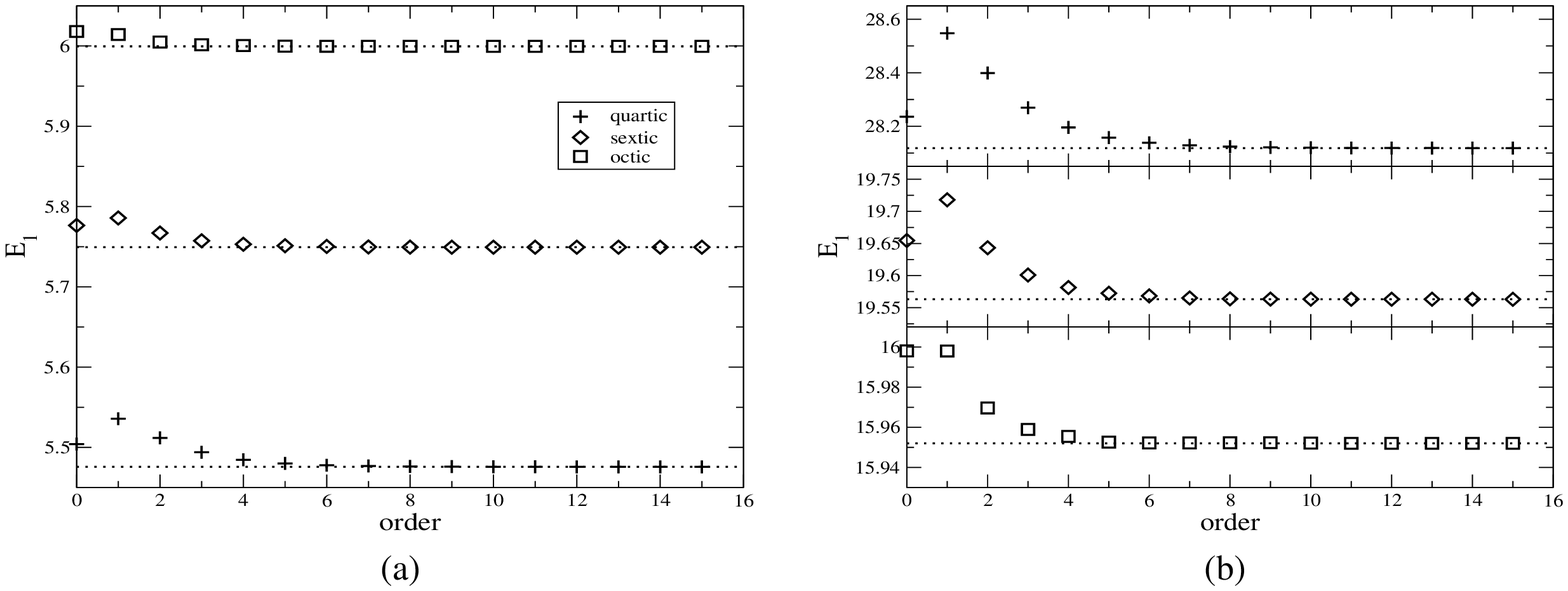}
\caption{Energy of the first excited state of the quartic, sextic and octic
  anharmonic oscillators, as a function of the approximation order, compared
  with the results of \protect\cite{mei97} (dotted). 
  We assume $m=1/2$ and $\omega=2$; $\mu= (8,12,16)$, respectively, in panel (a)
  and $\mu= (1600,2400,3200)$, respectively, in panel (b).} 
\label{fig:E1meis}
\vskip 3cm
\includegraphics[width=0.95\textwidth]{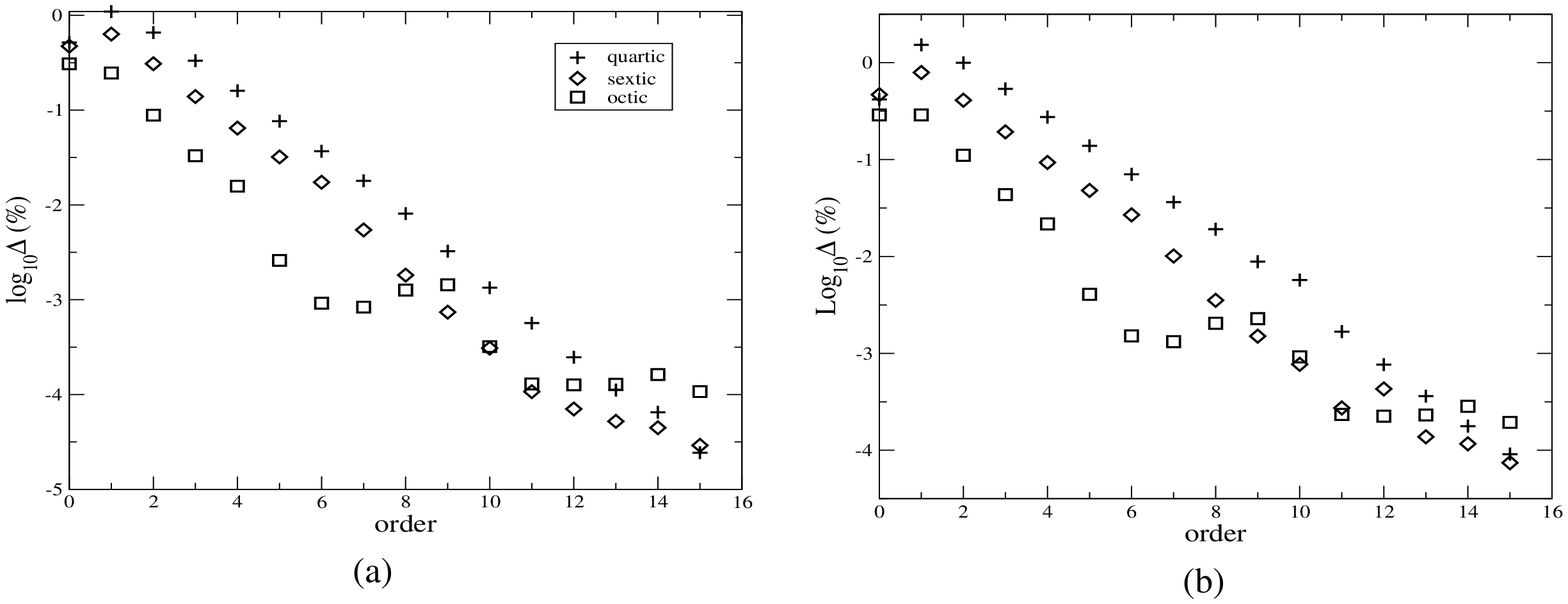}
\caption{Logarithm of the error, with respect to the results of
  \protect\cite{mei97}, of the energy of the first excited state of
  the quartic, sextic and octic anharmonic oscillators, as a function of the
  approximation order. 
  We assume $m=1/2$ and $\omega=2$; $\mu= (8,12,16)$, respectively, in panel (a)
  and $\mu= (1600,2400,3200)$, respectively, in panel (b).}
\label{fig:D1meis}
\end{center}
\end{figure}
The same considerations apply also to the excited states of the oscillators we
are studying, as can be inferred from Figs.~\ref{fig:E1meis} and
\ref{fig:D1meis}, which display the same quantities (energy and error) of
Figs.~\ref{fig:E0meis} and \ref{fig:D0meis}, but for the first excited state.

\begin{table}
\caption{\label{tab:meisbeta2} Energies of the ground state and of the first
  excited state calculated with our method to order $15$ and compared to the
  case $\beta=2$ in Table~III of \protect\cite{mei97}. 
  The correct digits are underlined.} 
\begin{ruledtabular}
\begin{tabular}{llll}
& quartic & sextic & octic  \\
\hline
$E_0$  & \underline{1.6075413}  & \underline{1.6099319}  &
\underline{1.64137}13  \\ 
$E_0$ (Ref.\cite{mei97})  & 1.607541302 & 1.609931952& 1.6413703 \\
\hline
$E_1$   &  \underline{5.47578}59 & \underline{5.74934}94 &
\underline{5.9996}138 \\ 
$E_1$  (Ref.\cite{mei97})  & 5.475784536  & 5.749347753 & 5.999607360 \\
\end{tabular}
\end{ruledtabular}
\caption{\label{tab:meisbeta400} As in Table~\protect\ref{tab:meisbeta2}, but
  for the case $\beta=400$ in Table~III of \protect\cite{mei97}. The correct
  digits are underlined.}  
\begin{ruledtabular}
\begin{tabular}{llll}
& quartic & sextic & octic  \\
\hline
$E_0$  & \underline{7.861862}8  & \underline{5.18835}9  & \underline{4.1461}938  \\ 
$E_0$ (Ref.\cite{mei97})  & 7.8618627 & 5.1883589 & 4.1461886 \\
\hline
$E_1$   &  \underline{28.1184}79 & \underline{19.5631}45 & \underline{15.95}2016 \\ 
$E_1$  (Ref.\cite{mei97})  & 28.118454  & 19.563130 & 15.951985 \\
\end{tabular}
\end{ruledtabular}
\end{table}
In Tables~\ref{tab:meisbeta2} and \ref{tab:meisbeta400} we report the actual
values of the energies of the ground state and of the first excited state
calculated with our method to order $15$ and compared to the results of
Ref.~\cite{mei97}, with AHO parameters corresponding to the cases $\beta=2$ and
$\beta=400$ of that reference, respectively.

\begin{figure}[t]
\begin{center}
\includegraphics[width=0.95\textwidth]{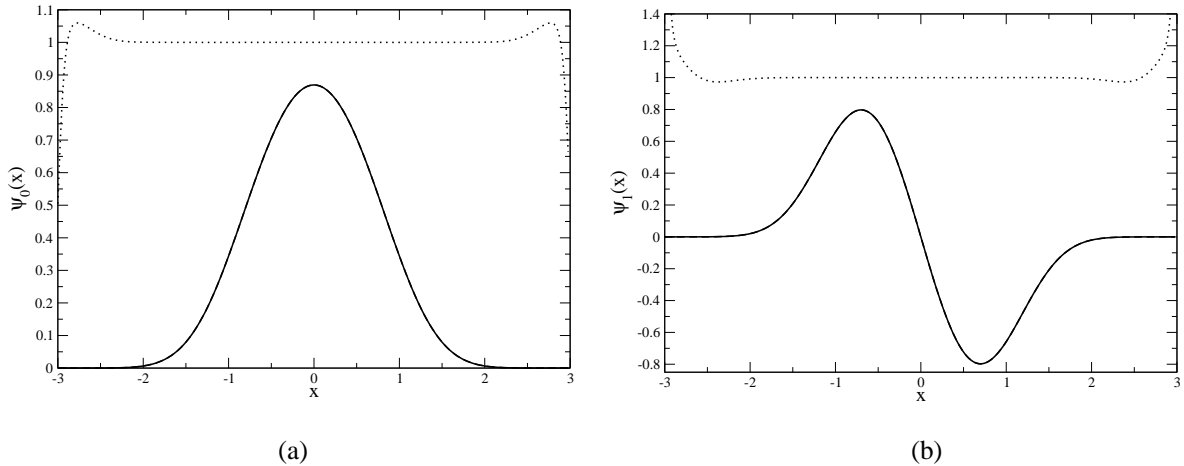}
\caption{Wave function for the ground state (a) and the first excited state (b)
  of the quartic anharmonic oscillator. The solid line is the numerical result,
  the dashed line (not visible) is the approximation to order $15$ of our
  method. The dotted line show the ratio $R(x) =
  \psi_i^{\text{(appr)}}(x)/\psi_i^{\text{(num)}}(x)$ of the
  approximate to the numerical wave function.  We assume $m=1/2$, $\omega=2$
  and $\mu= 8$.}
\label{fig:wavefunct}
\end{center}
\end{figure}

Good accuracy is also obtained for the wave functions. 
In Fig.~\ref{fig:wavefunct} one can see the ground state and the first excited
state wave functions of the quartic AHO as obtained in our approximation and by
direct numerical calculation. 

The numerical calculation was performed using a Fortran program. We have tested
the accuracy of the program by calculating the wave function of the harmonic 
oscillator and comparing it to the exact result. The error defined as
$\Delta_\psi(x) = 100 \times
\left|(\psi^{(\text{num})}(x)-\psi^{(\text{exact})}(x))/\psi^{(exact)}(x)
\right|$ is found to be smaller than $10^{-7}$ in the region where the wave
function is sizable. 
As a result, a meaningful comparison of our approximate results with the
numerical results is possible.

Actually, the two curves in Fig.~\ref{fig:wavefunct} are not distinguishable
from each other at the scale of the figures: For this reason we display also
the ratio of the approximate to the exact wave function. 

\section{Conclusions}
\label{sec_conc}

In Ref.~\cite{AAD03} a new method for the solution of the Schr\"odinger equation was 
introduced. It is based on the application of the linear delta expansion (optimized
perturbation theory) and an ansatz for the wave function that explicitly takes into account
its asymptotic behavior. The new method was applied to calculate the energies and wave 
functions of the ground and first excited state of the quartic anharmonic potential. 

In this paper we have extended the results of Ref.~\cite{AAD03} in two ways, first we have
computed the energies by evaluating the expectation value of the Hamiltonian at a given order
using the wave function obtained with the method and shown that the accuracy obtained is greater.
Secondly, we have extended the results by computing the energies and wave functions of the
ground and first excited states for the quartic, sextic and octic potentials. We have verified
that the method works very well for these potentials and that in fact one is able to obtain high
accuracy with only a few perturbative orders. We have also presented quantitative comparisons 
between our results and other methods found in the literature.

\begin{acknowledgments}
P.A. and A.A. acknowledge support for this work to the ``Fondo Alvarez-Buylla''
of Colima University. P.A. also acknowledges Conacyt grant no. C01-40633/A-1.
J.A.L. thanks the warm hospitality of the Universidad de Colima while this 
work was in progress.

\end{acknowledgments}

\end{document}